# Charmonium spectroscopy with heavy Kogut-Susskind quarks *

S. Aoki[a], M. Fukugita[b], S. Hashimoto[c, d], N. Ishizuka[a, e], H. Mino[f], M. Okawa[d], T. Onogi[c, g] and A. Ukawa[a]

[a]Institute of Physics, University of Tsukuba, Tsukuba, Ibaraki 305, Japan

[b]Yukawa Institute for Theoretical Physics, Kyoto University, Kyoto 606, Japan

[c]Department of Physics, Hiroshima University, Higashi-Hiroshima 724, Japan

[d]National Laboratory for High Energy Physics(KEK), Tsukuba, Ibaraki 305, Japan

[e]Department of Physics, Washington University, St. Louis, MO 63130-4899, USA

[f]Faculty of Engineering, Yamanashi University, Kofu 400, Japan

[g]Theory Group, Fermi National Accelerator Laboratory, Batavia, IL 60510, USA

Charmonium spectroscopy with Kogut-Susskind valence quarks are carried out for quenched QCD at $\beta = 6.0$ and for two-flavor full QCD at $\beta = 5.7$. Results for 1P–1S mass splitting and estimates of $\alpha^{(5)}_{\overline{MS}}(m_Z)$ are reported. Problems associated with flavor breaking effects and finite size effects of $1P$ states are discussed.

## 1. INTRODUCTION

An interesting application of lattice heavy quark QCD is a determination of the strong coupling constant through the 1P–1S mass splitting. A shortcoming of the initial studies[1], however, was the use of quenched simulations, necessitating an estimate of sea quark effects via the perturbative renormalization group. While a mean-field estimate of the renormalized coupling in quenched and full QCD provided support for the validity of this procedure[2], a direct quarkonium spectrum simulation in full QCD is clearly desirable. An exploratory calculation has been carried out in Ref.[2] using Wilson valence quarks on full QCD configurations generated with two flavors of Kogut-Susskind (KS) quarks. Here we report on our attempt to employ the KS action also for heavy valence quarks. Our study is carried out in two steps; a detailed quenched study of the charmonium spectrum for examining the viability of the KS action for heavy quarks in relation to flavor breaking effects, followed by an extension to full QCD and estimates of the strong coupling constant.

*presented by S. Aoki

Table 1
Run parameters

| | $\beta$ | $a^{-1}$(GeV) | size | #conf. | $m_h$ |
|---|---|---|---|---|---|
| quenched | 6.0 | 1.88(6) | $24^3 \times 40$ | 50 | 0.3,0.4,0.5 |
| | | | $16^3 \times 40$ | 60 | 0.3,0.4,0.5 |
| full QCD | 5.7 | 2.23(9) | $20^3 \times 20$ | 38 | 0.2,0.3,0.4 |

## 2. SIMULATION

Parameters of our runs are listed in Table 1 including the values of heavy quark mass $m_h a$ and the lattice spacing determined from $m_\rho$. For full QCD we use the configurations generated previously with the sea quark mass of $m_{sea}a = 0.01$ and 0.02[3], taken at 20 trajectory intervals and periodically doubled in time. Quenched configurations are those employed in Refs. [4,5].

We employ 128 meson operators extended over a $2^4$ hypercube and non-local in time. These are classified into 8 GTS irreducible representations for $\eta_c$, 12 for $J/\psi$, 6 for $h_c$, and 4 and 6 for $\chi_{c_0}$ and $\chi_{c_1}$[6]. Eight wall sources at $t = 0$ and 1 are used to maximize propagator signals[4] calculated with Landau gauge fixing. Hadron propagators are averaged over members of irreducible representations before fitting. Errors are estimated by the single elimination jackknife procedure.

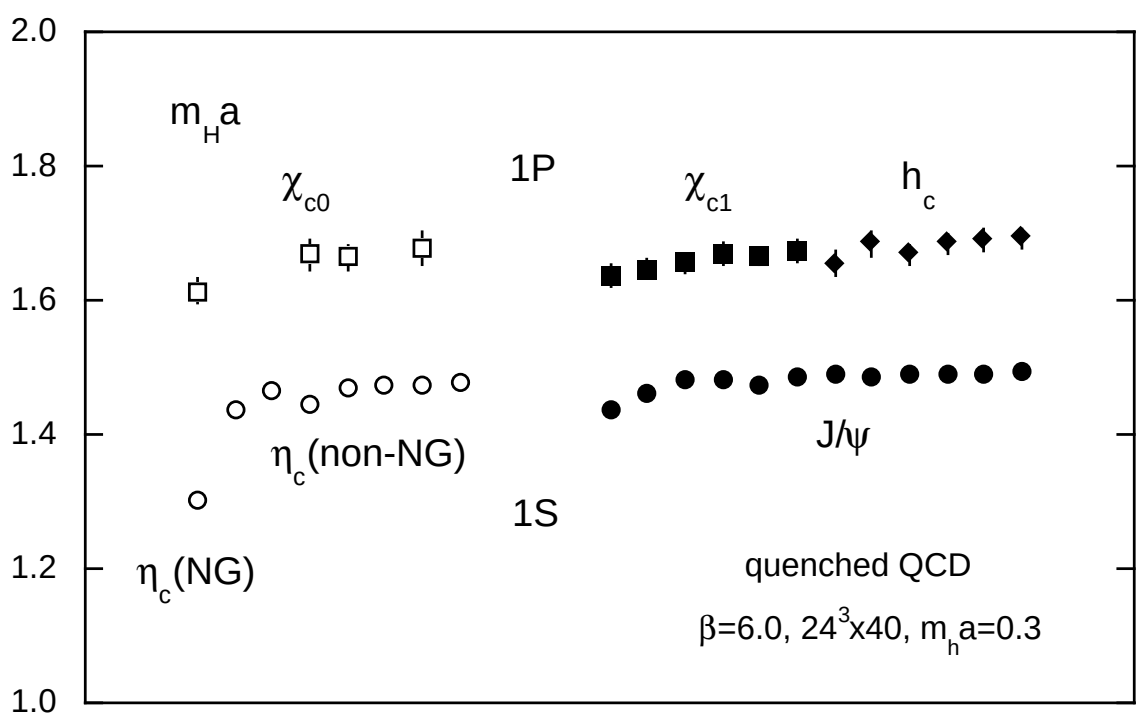

Figure 1. Operator dependence of quenched meson masses at $\beta$=6.0 and $m_h a$=0.3 on a $24^3 \times 40$ lattice.

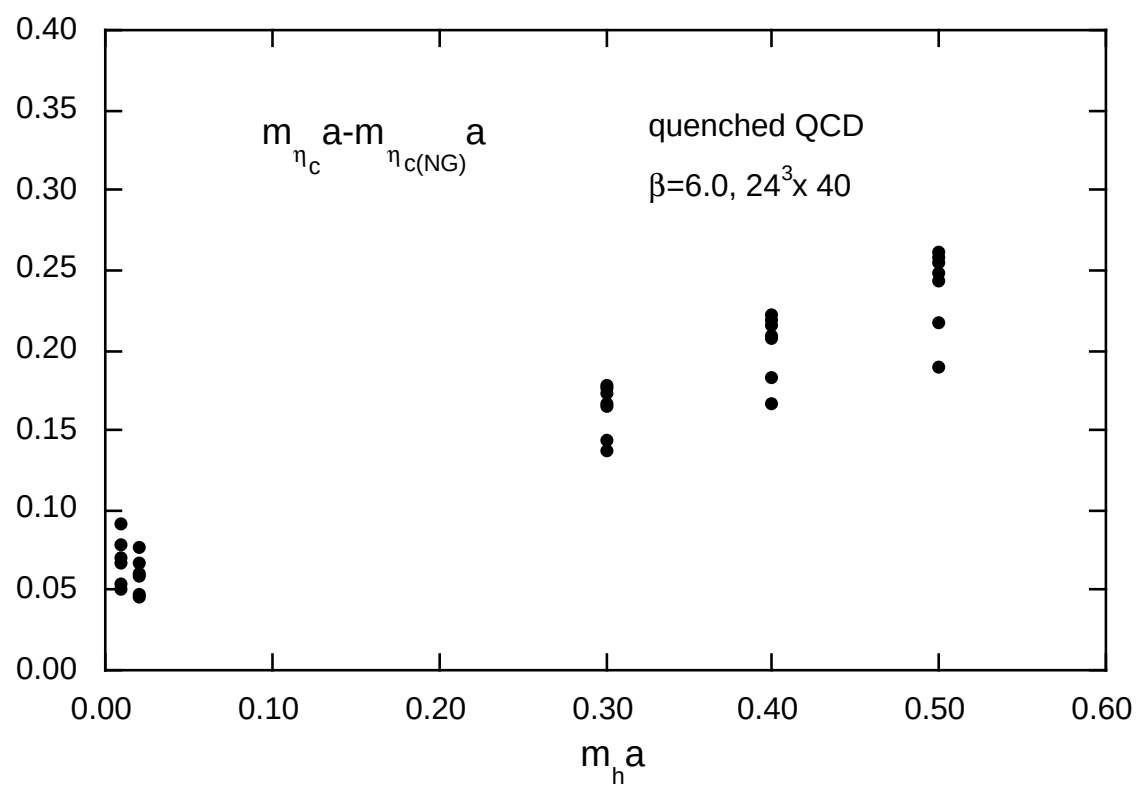

Figure 2. $m_{\eta_c} a - m_{\eta_c(NG)} a$ for 5 non-Nambu-Goldstone $\eta_c$ states as a function of $m_h a$ at $\beta$=6.0 on a $24^3 \times 40$ lattice in quenched QCD.

## 3. CHARMONIUM SPECTRUM

The operator dependence of charmonium masses is illustrated in Fig. 1. A notable feature is that $\eta_c(NG)$ in the Nambu-Goldstone channel is much lighter than the others ($\eta_c(non\text{-}NG)$) even at heavy quark masses. It is quite disturbing to find that the mass splitting increases for larger quark mass, as shown in Fig. 2 where the splitting relative to the Nambu-Goldstone pseudo scalar is plotted as a function of $m_h a$ for the 5 GTS irreducible representations for non-NG $\eta_c$. Other states also exhibit a similar increase of mass splitting with $m_h a$, though much smaller in magnitude.

There is no reason to favor a particular channel when comparing measured masses with the experiment. We, therefore, consider meson masses averaged over GTS representations having the same continuum $J^{PC}$ weighted by their dimensions. For $\eta_c$, however, we separate out $\eta_c(NG)$. In Fig. 3 we plot the averaged masses in physical units for quenched QCD employing the scale determined by $m_\rho$, together with the experimental values. For the states other than $\eta_c$ the calculation reproduces qualitative features of the charmonium spectrum. However, we find $m_{J/\psi} - m_{\eta_c(NG)} = 404(4)$MeV, which is much larger than the experimental value 118MeV, while a factor two smaller value of 51(6)MeV is obtained with the average over non-NG $\eta_c$.

Results for the charmonium spectrum for full QCD is quantitatively similar to the quenched results. For $J/\psi - \eta_c$ mass splitting, in particular, we find 442(4)MeV for $\eta_c(NG)$ and 62(7)MeV for $\eta_c(non\text{-}NG)$. A 15% difference in the lattice spacing between our full and quenched simulations makes it difficult to see if the larger splitting for full QCD represents a physical effect of sea quarks that is expected from renormalization group considerations.

## 4. 1P–1S MASS SPLITTING

Our results for the charmonium 1P-1S mass splitting for quenched and full QCD are summarized in Fig. 4. For quenched QCD the splitting is almost independent of $m_h a$ for each lattice size, which is consistent with the experiment. There exists, however, a 20% difference between the values obtained with $\eta_c(NG)$ and $\eta_c(non\text{-}NG)$.

Another important point is a significant lattice size dependence of the splitting between $L = 16$ and 24 observed for quenched QCD. We found that this arises from a size dependent mass of the $1P$ state $h_c$, which increases by 2.4(9)% between the two lattice sizes (for $J/\psi$ and $\eta_c$ the change is less than 0.2 %). Though small for the mass itself, this shift translates into a 20 % increase for the 1P–1S splitting.

The physical lattice size of $La \simeq 1.8$fm for full QCD on an $L = 20$ lattice is comparable to $La \simeq$1.7fm for quenched QCD on an $L = 16$ lattice. In view of the size dependence for the quenched result it is likely that the full QCD value for the 1P–1S mass splitting is an underestimate. In fact, employing the experimental value $\Delta m_{1P-1S}$ = 458MeV, we find $a^{-1}$=2.70(20)GeV with $\eta_c(NG)$ and 3.59(36)GeV with $\eta_c(non\text{-}NG)$ at $m_{sea}a$=0.01 and $m_h a$=0.3, which are significantly larger than the value 2.23(9)GeV estimated from $m_\rho$. The corresponding values for the quenched case on an $L$ = 24 lattice are 1.87(16) GeV with $\eta_c(NG)$ and 2.32(25)GeV with $\eta_c(non\text{-}NG)$ at $m_h a$=0.4, which are compared with 1.88(6)GeV from $m_\rho$.

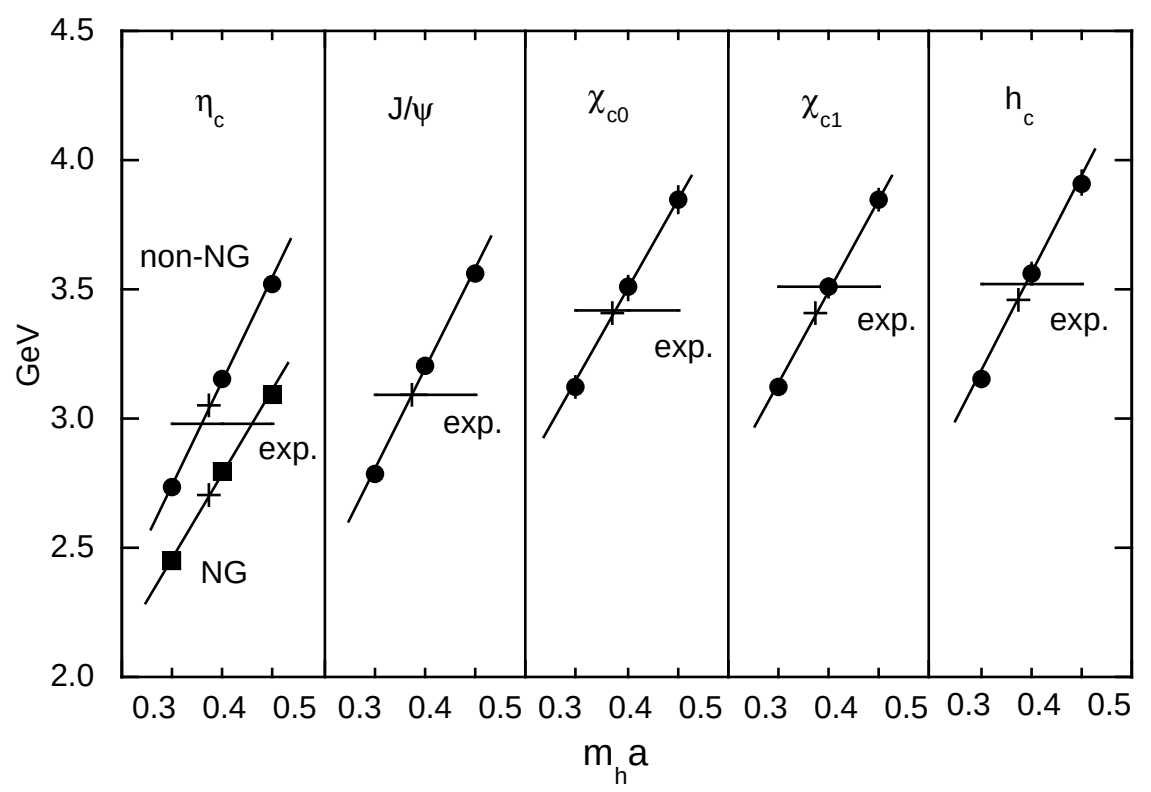

Figure 3. Averaged meson masses in physical units as functions of $m_h a$ at $\beta$=6.0 on a $24^3 \times 40$ lattice in quenched QCD with the scale fixed by $m_\rho$. Crosses correspond to the point where the mass of $J/\psi$ agrees with the experiment.

## 5. DETERMINATION OF $\alpha_{\overline{MS}}^{(5)}(m_Z)$

Our method for estimating the strong coupling constant follows that of Ref.[1]. The $\alpha_V$ coupling is used for renormalization group evolution and $\mu$ = 0.4 − 0.75 GeV is taken for the charmonium scale. For quenched QCD we calculate $\alpha_{\overline{MS}}^{(5)}(m_Z)$ for three values of $m_h a$ and the two choice of inverse lattice spacing corresponding to $\eta_c(NG)$ and $\eta_c(non\text{-}NG)$. Averaging the resulting six values of $\alpha_{\overline{MS}}^{(5)}(m_Z)$, we obtain $\alpha_{\overline{MS}}^{(5)}(m_Z)$ = 0.111(4)(4). The first error reflects the ambiguity of the charmonium matching scale and the second the choice of $\eta_c(NG)$ or $\eta_c(non\text{-}NG)$ for determining the scale. For full QCD a similar analysis yields $\alpha_{\overline{MS}}^{(5)}(m_Z)$ = 0.119(3)(3).

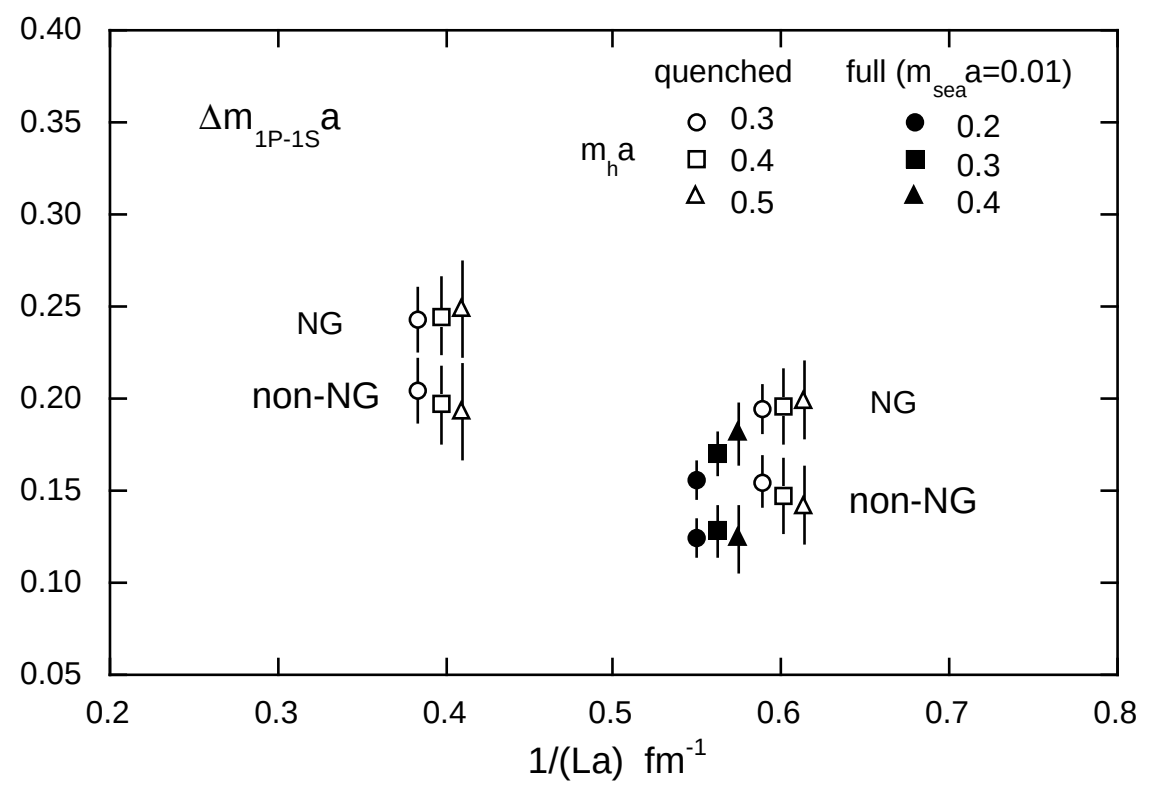

Figure 4. 1P–1S mass splittings in lattice units for quenched (open symbols) and full (filled symbols) QCD for various quark masses as a function of $1/La$ with the scale fixed by $m_\rho$.

The true value must be smaller than this due to an over-estimate of the lattice scale as discussed above.

## 6. Summary

Charmonium spectroscopy with the KS action suffers from a sizable flavor breaking effect whose magnitude increases for heavier quarks, particularly for $\eta_c$, at a currently used lattice spacing. Systematic uncertainties in $\alpha_{\overline{MS}}^{(5)}(m_Z)$ due to this source is difficult to be diminished. Whether modification of the action is possible for reducing the effect is an open problem at present.